\documentclass[twocolumn,PRA,preprintnumbers,superscriptaddress,amsmath]{revtex4}
\usepackage[latin9]{inputenc}
\usepackage{xcolor}
\usepackage{pdfcolmk}
\usepackage{amsmath}
\usepackage{amssymb}
\usepackage{graphicx}
\PassOptionsToPackage{normalem}{ulem}
\usepackage{ulem}

\makeatletter

\providecolor{lyxadded}{rgb}{0,0,1}
\providecolor{lyxdeleted}{rgb}{1,0,0}

\@ifundefined{textcolor}{}
{%
 \definecolor{BLACK}{gray}{0}
 \definecolor{WHITE}{gray}{1}
 \definecolor{RED}{rgb}{1,0,0}
 \definecolor{GREEN}{rgb}{0,1,0}
 \definecolor{BLUE}{rgb}{0,0,1}
 \definecolor{CYAN}{cmyk}{1,0,0,0}
 \definecolor{MAGENTA}{cmyk}{0,1,0,0}
 \definecolor{YELLOW}{cmyk}{0,0,1,0}
}

\makeatother

\begin{document}

\title{Non-reciprocal nonlinear optic induced transparency and frequency
conversion on a chip}

\author{Xiang Guo}

\address{Department of Electrical Engineering, Yale University, New Haven,
Connecticut 06511, USA}

\author{Chang-Ling Zou}

\address{Department of Electrical Engineering, Yale University, New Haven,
Connecticut 06511, USA}

\author{Hojoong Jung}

\address{Department of Electrical Engineering, Yale University, New Haven,
Connecticut 06511, USA}

\author{Hong X. Tang{*}}

\address{Department of Electrical Engineering, Yale University, New Haven,
Connecticut 06511, USA}

\maketitle
\textbf{Developments in photonic chips have spurred photon based classical
and quantum information processing, attributing to the high stability
and scalability of integrated photonic devices \cite{Welch_IEEE,Khilo-OE}.
Optical nonlinearity \cite{Shen-Book} is indispensable in these complex
photonic circuits, because it allows for classical and quantum light
sources, all-optical switch, modulation, and non-reciprocity in ambient
environments. It is commonly known that nonlinear interactions are
often greatly enhanced in the microcavities \cite{Vahala-Nature}.
However, the manifestations of coherent photon-photon interaction
in a cavity, analogous to the electromagnetically induced transparency
\cite{Fleischhauer-RMP}, have never been reported on an integrated
platform. Here, we present an experimental demonstration of the coherent
photon-photon interaction induced by second order optical nonlinearity
($\chi^{(2)}$) on an aluminum nitride photonic chip. The non-reciprocal
nonlinear optic induced transparency is demonstrated as a result of
the coherent interference between photons with different colors: ones
in the visible wavelength band and ones in the telecom wavelength
band. Furthermore, a wide-band frequency conversion with an almost
unit internal ($0.14$ external) efficiency and a bandwidth up to
$0.76\,\mathrm{GHz}$ is demonstrated.}

The importance of integrating nonlinear devices on a photonic chip
has become more prominent due to the devices' small foot-prints and
large scalability \cite{Moss-NO,Hausmann-NO}. Second order optical
nonlinearity ($\chi^{(2)}$) is one of the most widely explored properties
in photonics, utilizing various nonlinear materials \cite{Rakher-NO,Xiong-OE,Levy-OE,De Greve-Nature,Xiong-NJP,Kuo-NC}.
$\chi^{(2)}$ nonlinearity enables the coupling between photons with
very different colors, acting as the basis for many important applications
such as second harmonic generation, spontaneous parametric down conversion,
optical parametric amplification and oscillation. Due to the high
quality factor to mode volume ratio, the nonlinear interaction strength
is expected to be boosted in optical cavities. Preliminary results
in millimeter sized optical resonators have already shown such trend,
where efficient second harmonic generation \cite{Ilchenko-PRL,Furst-PRL}
and sum frequency generation \cite{Strekalov-NJP} are demonstrated.
However, the realized nonlinear interaction strength on an integrated
platform is normally weak, hindered by the challenges of fabricating
small size, low loss optical circuits with materials featuring high
$\chi^{(2)}$ nonlinearity. 

In this Letter, we demonstrate coherent interaction between photons
of different colors on a scalable aluminum nitride-on-insulator \cite{Xiong-NJP}
chip based on $\chi^{(2)}$ optical nonlinearity. The nonlinear optic
induced transparency (NOIT), as an analogue to electromagnetically
induced transparency resulting from coherent photon-atom \cite{Fleischhauer-RMP}
or photon-phonon interactions \cite{Weis-Science,Safavi-Naeini-Nature,Kim-NP,Dong-NC},
is reported. Due to the inherent phase matching condition, the $\chi^{(2)}$
nonlinearity based coherent interaction and the accompanying NOIT
phenomenon are non-reciprocal \cite{Kim-NP,Dong-NC}, which permits
future applications such as non-magnetic, ultrafast optical isolators
\cite{Jalas-NO,Tzuang-NO,Peng-NP}. We further realize a saturated
frequency conversion from telecom band to visible band with almost
unit internal ($0.14$ external) conversion efficiency. Notably, the
transparency window demonstrated here is close to $\mathrm{GHz}$,
which is orders of magnitudes larger than the bandwidth of previously
demonstrated transparency \cite{Weis-Science,Safavi-Naeini-Nature,Kim-NP,Dong-NC}
and wavelength conversion \cite{Dong-Science,Hill-NC} induced by
opto-mechanical interaction. This large bandwidth, together with a
flexible working environment (ambient air - no need for cryogenic
cooling or vacuum system), may be beneficial for applications including
optical switches, isolators and frequency converters.
\begin{figure*}[tp]
\includegraphics[width=17cm]{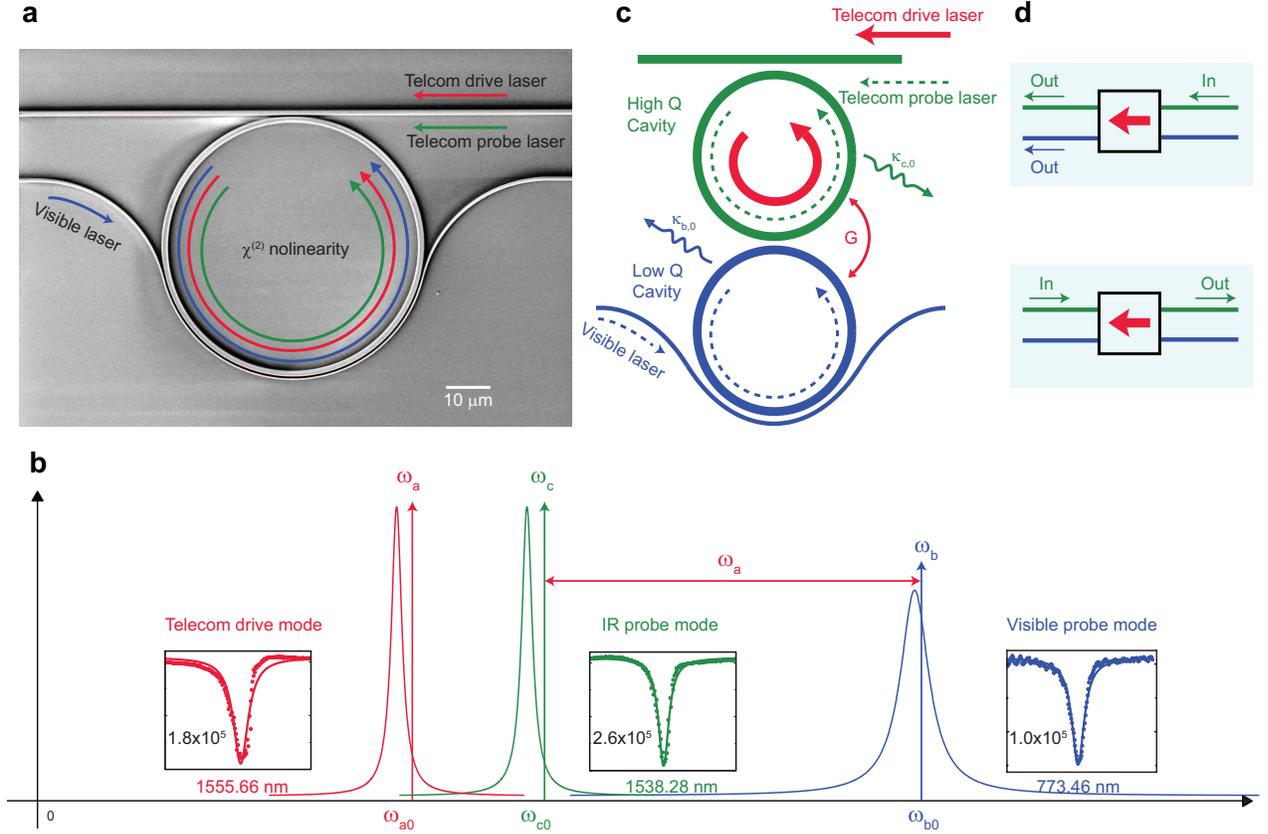}

\protect\caption{\textbf{Triply resonant microring resonators coupled by second order
optical nonlinearity. a}, SEM picture of the core device. Three transverse-magnetic
modes (two in the telecom band, one in the visible band) co-exist
in the microring resonator, and couple through $\chi^{(2)}$ nonlinear
interaction. \textbf{b}, Schematic sketch of the density of states
for the three modes and their respective transmission spectrum. At
the triply resonant condition, the three modes' frequencies approximately
fulfill the energy conservation requirement $\omega_{a0}+\omega_{c0}=\omega_{b0}$.
\textbf{c},\textbf{ }Coupled resonator model for the coherent interaction
in a $\chi^{(2)}$ microring resonator. Here we use two resonators
to depict resonant modes supported at two different wavelengths. The
interaction between them is through $\chi^{(2)}$ nonlinearity and
the coupling strength is controlled by the power of the drive laser.\textbf{
d}, A further simplified four-port block model to describe the nonlinear
interaction. Here the device works as a two-in-two-out beamsplitter,
whose splitting ratio is controlled by the power of the drive laser.
The device is non-reciprocal due to the required momentum conservation
condition in $\chi^{(2)}$ conversion process. }
\label{Fig1}
\end{figure*}

Figure 1a illustrates the fabricated aluminum nitride ($\mathrm{AlN}$)
microring structure, which includes two bus waveguides that couple
with the telecom and visible modes in the microring separately. The
$\chi^{(2)}$ nonlinearity of the $\mathrm{AlN}$ induces the interaction
between three non-degenerate modes, as described by the Hamiltonian
\begin{equation}
\mathcal{H}=\omega_{a0}\hat{a}^{\dagger}\hat{a}+\omega_{b0}\hat{b}^{\dagger}\hat{b}+\omega_{c0}\hat{c}^{\dagger}\hat{c}+g(\hat{a}\hat{b}^{\dagger}\hat{c}+\hat{a}^{\dagger}\hat{b}\hat{c}^{\dagger}).
\end{equation}
Here, $\hat{a}$, $\hat{b}$ and $\hat{c}$ are the bosonic operators
for three transverse-magnetic (TM) modes in the microring. In our
experiment, we choose modes $a$, $c$ in the telecom band and mode
$b$ in the visible band. On one hand, the momentum conservation condition
($m_{a}+m_{c}=m_{b}$) is required for a non-vanishing interaction
strength $g$, which demands modes traveling in the same direction
in the microring resonator. On the other hand, the nonlinear interaction
in the microring can only be maximally enhanced when the energy conservation
condition $\omega_{a0}+\omega_{c0}=\omega_{b0}$ is satisfied. We
carefully engineer the device geometry and find the modes that satisfy
both conservation conditions. The measured transmission spectra of
the modes (with loaded quality factors $Q_{a,b,c}=(1.8,1.0,2.6)\times10^{5}$)
and the schematic diagram of their density of states in frequency
domain are presented in Fig.$\,$1b. 

As we aim for the coherent conversion between the telecom and visible
photons, we strongly drive mode $a$ by a near resonance strong laser
($\omega_{a}$) and hence stimulate large exchange coupling strength
between mode $b$ and $c$. The simplified system Hamiltonian reads
\begin{equation}
\mathcal{H}=\omega_{b0}\hat{b}^{\dagger}\hat{b}+\omega_{c0}\hat{c}^{\dagger}\hat{c}+G\hat{c}\hat{b}^{\dagger}+G^{*}\hat{c}^{\dagger}\hat{b},
\end{equation}
where $G=\left\langle \hat{a}\right\rangle g$ is the effective coupling
strength, $\left|\left\langle \hat{a}\right\rangle \right|^{2}\propto P_{a}$
is the mean photon number of mode $a$ and $P_{a}$ is the power of
the drive laser. This beamsplitter-like Hamiltonian indicates that
photons in cavities $b$ and $c$ can be converted to each other coherently
without introducing additional noises \cite{Huang-PRL}, just like
a linear optics device.

It is instructive to introduce a simplified model, as depicted in
Fig.$\,$1c, where two resonators loaded by two separate bus waveguides
are used to represent the resonant modes $b$ and $c$ separately.
The originally independent resonators are coupled together through
nonlinear interaction $G$, which is enabled and controlled by drive
laser $\omega_{a}$. Intuitively the photon flux input from top waveguide
can be dropped to the bottom waveguide, or vice versa, analogous to
the coupled-microring add-drop filter \cite{Little-JLT}. Such a device
can be schematically illustrated as a four-port block (Fig.$\,$1d),
exhibiting two intriguing properties: first, the color of dropped
photon changes due to nonlinear wavelength conversion; second, the
nonlinear interaction process is non-reciprocal because $G$ vanishes
for the probe lights propagating in the opposite direction of the
drive laser due to momentum mismatch. 

\begin{figure*}[tp]
\includegraphics[width=17cm]{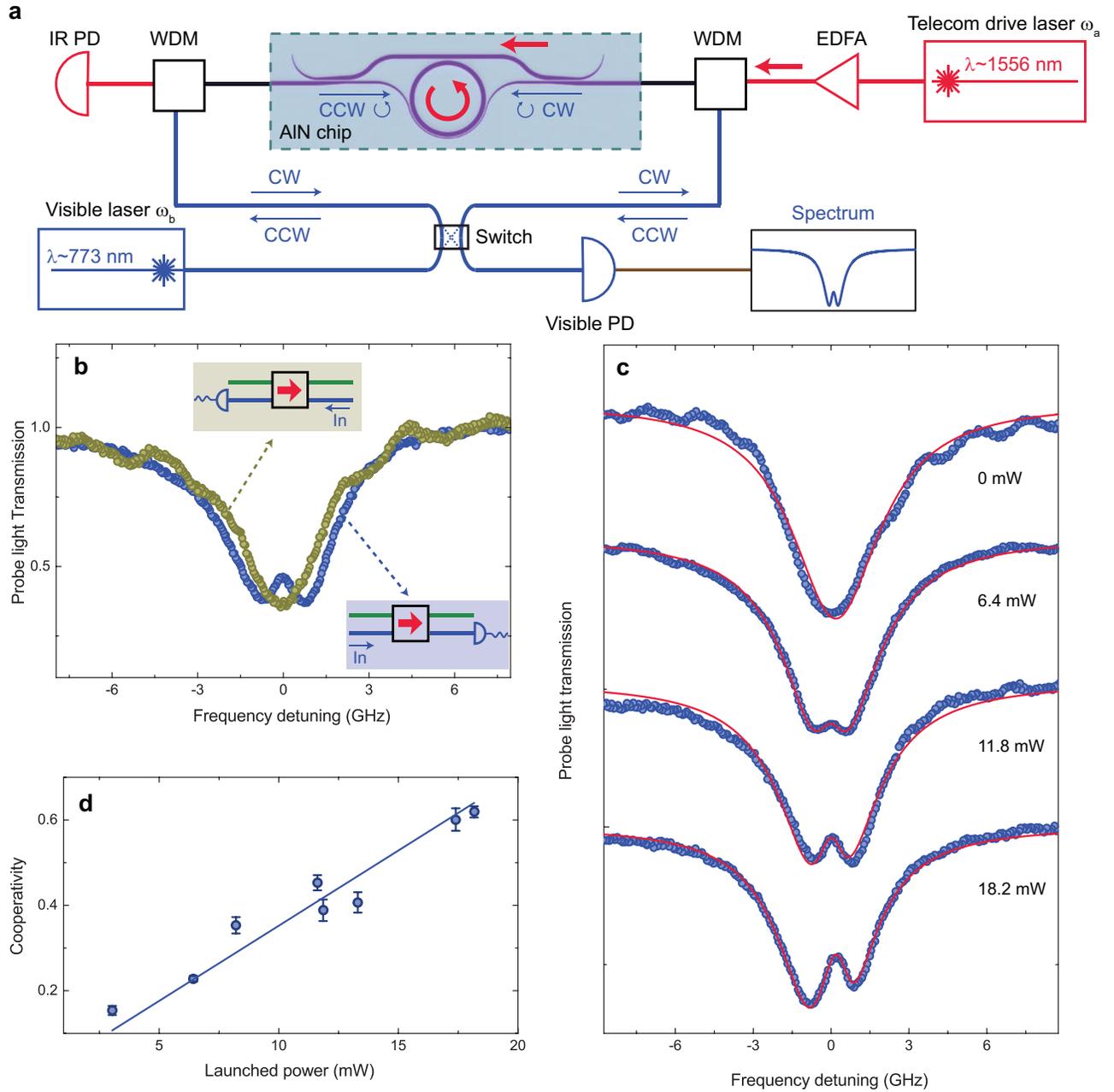}

\protect\caption{\textbf{Nonlinear optic induced transparency. a}, Experimental setup.
After amplification with an EDFA, telecom laser $\omega_{a}$ excites
mode $a$ in the microring resonator with counter-clockwise (CCW)
propagation direction. A visible probe laser $\omega_{b}$ is used
to measure the modified transmission spectrum of mode $b$ (centered
at $\omega_{b0}$). To verify non-reciprocity, the probe laser's propagation
direction is switchable. \textbf{b}, Transmission spectra of mode
$b$ with both CW and CCW propagation directions. A transparency window
is observed if the probe laser propagates in the same direction (CCW)
as the drive laser. When the probe laser propagates in the other direction
(CW in this case), no transparency window is observed. \textbf{c},\textbf{
}NOIT spectra under different powers of the drive laser. The transparency
peak's height and width increase with the drive powers, which are
labeled in the figure. The red solid lines are fittings with the theoretical
transmission formula shown in Eq. \ref{eq:Pb}. The cooperativities
can be extracted from the fittings. \textbf{d}, Extracted cooperativities
at different drive laser powers. A maximum cooperativity of $0.62$
is achieved. A linear dependence is observed with an unit power cooperativity
to be $0.035\pm0.001\,\mathrm{(1/mW)}$.}
\label{Fig2}
\end{figure*}
\begin{figure*}[!tp]
\includegraphics[width=17cm]{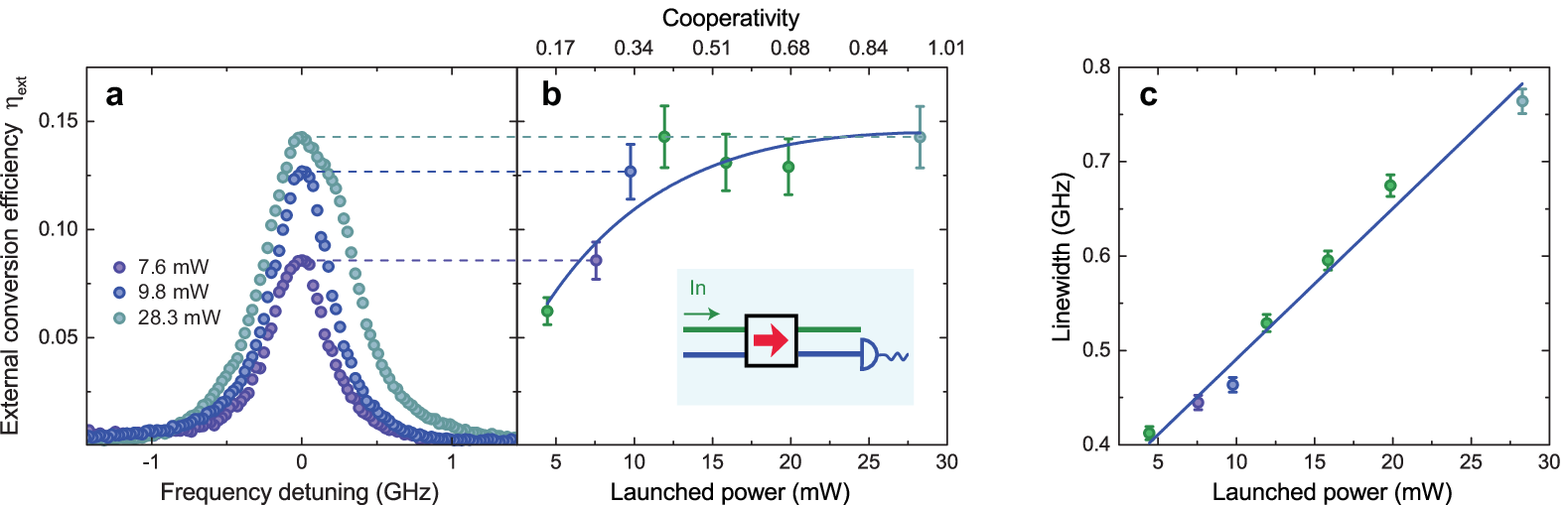}

\protect\caption{\textbf{Optical frequency conversion from the telecom band to the
visible band. a}, Frequency conversion spectra with telecom probe
laser $\omega_{c}$ tuned across the resonance of mode $c$ under
different drive laser powers, which are labeled in the figure. \textbf{b},
Maximum external conversion efficiency against different drive power.
The error bar comes from the uncertainty of the fiber-to-chip insertion
loss. The blue solid line is a fitted curve according to Eq. (\ref{eq:eta_max}).
A unit power cooperativity of $0.034\pm0.006\,\mathrm{mW^{-1}}$ is
deduced., which is consistent with the values measured in NOIT experiment.
Inset, the four-port block model for the frequency conversion process.\textbf{
c},\textbf{ }The linewidth of the frequency conversion spectra. The
error bar comes from the fitting error of each frequency conversion
spectrum. A broadened linewidth is observed with the increase of the
drive laser power, which results from Purcell effect.}
\label{Fig3}
\end{figure*}

From the device schematic illustrated by Fig.$\,$1c, the most straightforward
outcome of the coherent coupling between two resonators is the modified
resonance spectrum. When the coupling strength $G$ is comparable
or larger than the dissipation rates of the resonators, photons can
cycle between the two resonances back and forth coherently and interference
can be introduced. The effective frequency and dissipation of the
resonance are hence changed due to backactions. We probe such effect
using the experimental setup shown in Fig. 2a. The telecom drive laser
($\omega_{a}$, near the resonant frequency $\omega_{a0}$ of mode
$a$) is amplified with an erbium doped fiber amplifier (EDFA) and
excites the counter-clockwise (CCW) propagating mode in the microring,
while a visible laser ($\omega_{b}$) probes the transmission of mode
$b$ (centered at $\omega_{b0}$) in either clockwise (CW) or CCW
directions, as controlled by an optical switch. Figure 2b plots the
spectra of mode $b$ with drive laser ($P_{a}=13.3\,\mathrm{mW}$)
fixed on the resonance of mode $a$. A transparency window in the
center of the optical resonance is observed when the visible laser
propagates in the same direction (CCW) as the drive laser. In contrast,
when the probe laser propagates in the counter-propagating direction
(CW) of the drive laser, the resonance is of normal Lorentzian shape,
similar to the transmission spectrum without parametric pumping (Fig.
2c, top panel). The comparison of the spectra for two propagating
directions reveals the non-reciprocal nonlinear effect in the microring,
and also unambiguously demonstrates the coherent interaction in a
$\chi^{(2)}$ cavity, which manifests as nonlinear optic induced transparency
(NOIT). 

The observed NOIT is further tested with different powers of the drive
laser (Fig.$\,$2c). A clear increase of transparency peak is observed
with the increase of launched drive laser power. Following the Hamiltonian
(Eq.$\,$2) and considering the linear losses of modes $b$ and $c$
($\kappa_{x,i}$, $x=b,c$ for two modes and $i=0,1$ for intrinsic
and external losses), we derive the theoretical formula for the transmission
of mode $b$ as 
\begin{equation}
T=\left|1+\frac{2\kappa_{b,1}}{-i\delta_{b}-\kappa_{b}+\frac{|G|^{2}}{-i\delta_{c}-\kappa_{c}}}\right|^{2},\label{eq:Pb}
\end{equation}
where $\kappa_{a(b)}=\kappa_{a(b),0}+\kappa_{a(b),1}$ is total loss
rate, $\delta_{b}=\omega_{b0}-\omega_{b}$ and $\delta_{c}=\omega_{c0}-(\omega_{b}-\omega_{a})$
are the angular frequency detunings of mode $b$ and $c$, respectively.
The experimental results are fitted according to Eq. (\ref{eq:Pb})
and exhibit valid agreements, as shown by the red solid lines in Fig.$\,$2c.
To quantify the strength of coherent nonlinear interaction, the cooperativity
$C=\frac{\left|G\right|^{2}}{\kappa_{b}\kappa_{c}}$ is extracted
from each transmission spectrum and plotted against drive laser power
$P_{a}$ (Fig.$\,$2d). The linear dependence of $C$ over the launched
drive power is observed, as expected by theory that $C\propto|G|^{2}\propto P_{a}$.
An unit power cooperativity of $C/P_{a}=0.035\pm0.001\,\mathrm{mW^{-1}}$
is derived and the maximum cooperativity achieved here is $0.62$,
corresponding to $\left|G\right|=2\pi\times0.72\,\mathrm{GHz}$. Considering
that $\kappa_{b}=2\pi\times1.84\,\mathrm{GHz}$, $\kappa_{c}=2\pi\times0.46\,\mathrm{GHz}$,
the system is working inside the NOIT regime ($\kappa_{c}<\left|G\right|<\kappa_{b}$)
\cite{Zhang-PRL}.

While the NOIT is the hallmark of the coherent interaction, another
manifestation of the large $C$ is the high efficiency frequency conversion
as the photons drop to different ports (Fig.$\,$1d). The device acts
as a beamsplitter for different colors of light and the splitting
ratio is controlled by the drive laser. As schematically illustrated
in the inset of Fig.$\,$3b, the external conversion efficiency (ratio
of output visible photons' number collected by bottom waveguide to
the input telecom photons' number in the top waveguide) is measured.
By sweeping the telecom input laser frequency $\omega_{c}$ across
the resonance of mode $c$ and monitoring the power output near the
resonance of mode $b$ (see experimental setup in supplementary section
IV), we measure a Lorentzian-like power spectrum (Fig.$\,$3a). The
external conversion efficiency reads 
\begin{equation}
\eta_{ext}=\frac{\kappa_{b,1}}{\kappa_{b}}\frac{\kappa_{c,1}}{\kappa_{c}}\times\frac{4C}{\left|(1+i\frac{\delta_{b}}{\kappa_{b}})(1+i\frac{\delta_{c}}{\kappa_{c}})+C\right|^{2}}.
\end{equation}
For the cases that $C\lesssim1$, the maximum external conversion
efficiency can be obtained for near-resonance condition that $\delta_{c}\approx\delta_{b}\approx0$
\begin{equation}
\eta_{ext,max}\approx\frac{\kappa_{b,1}}{\kappa_{b}}\frac{\kappa_{c,1}}{\kappa_{c}}\times\frac{4C}{\left|1+C\right|^{2}}.\label{eq:eta_max}
\end{equation}
Figure 3b shows a saturated conversion efficiency against the launched
drive laser power. By fitting the experimental results with Eq. (\ref{eq:eta_max}),
we deduce an unit power cooperativity of $0.034\pm0.006\,\mathrm{mW^{-1}}$,
which agrees well with the value obtained from NOIT measurement. With
the pump power of $28.5\,\mathrm{mW}$, the highest cooperativity
we achieve is $0.97\pm0.17$. Accroding to Eq. (\ref{eq:eta_max}),
the saturated external conversion efficiency will be achieved when
$C\approx1$, which is measure to be $0.14$ in our experiment. In
terms of maximum internal conversion efficiency ($\eta_{int,max}=\frac{4C}{\left|1+C\right|^{2}}$)
\cite{Hill-NC}, which ignores the loss due to non-ideal waveguide-cavity
coupling ($\frac{\kappa_{b,1}}{\kappa_{b}}\frac{\kappa_{c,1}}{\kappa_{c}}<1$),
the conversion efficiency is $0.988\leq\eta_{int,max}\leq1$ as deduced
from the cooperativity of $0.97\pm0.17$. We further investigate the
dependence of the bandwidth of the frequency conversion on $P_{a}$,
which is shown in Fig.$\,$3c. Owing to a much larger linewidth of
mode $b$ compared to that of mode $c$, the bandwidth of frequency
conversion is broadened for large cooperativity $C$ due to the Purcell
effect \cite{Zhang-PRL}. We get a largest bandwidth of $0.76\,\mathrm{GHz}$
with a drive laser power of $28.5\,\mathrm{mW}$.

The cooperativity of the system may be further improved by more sophisticated
cavity design to reduce the cavity mode volume while maintaining the
high quality factors. Alternatively, by applying high peak power ultrafast
driving pulses, the instantaneous $C$ may be improved by 1 or 2 orders
of magnitude. The potential applications of the demonstrated non-reciprocal
NOIT are all-optical isolators and switches, with the modulation bandwidth
exceeding $\mathrm{GHz}$. The external efficiency and bandwidth of
the frequency conversion can also be improved, by increasing the external
coupling rate of the microring cavity while keeping the cooperativity
at $1$. For example, if the drive mode photon number increases by
$10$ times, the maximum achievable external conversion efficiency
is about $70\%$ and the bandwidth is about $2\,\mathrm{GHz}$ for
current experimental parameters. Such devices can be key components
for connecting matter qubits at visible wavelength with flying qubits
at telecom wavelength, enabling the distributed quantum computation
network and quantum communications \cite{Kimble-Nature,Raymer-PT}.

\vbox{}

\noindent\textbf{\large{}Methods}{\large \par}

\noindent\textbf{Device fabrication and design parameters. }The devices
are fabricated using AlN-on-oxide-on-Si chip. After defining the pattern
using electronic beam lithography, the waveguide and microring resonators
are dry etched using $\mathrm{Cl_{2}/BCl_{3}/Ar}$ chemistry. We then
deposit $\mathrm{2.5\,\mu m}$ thick PECVD oxide on top of the $\mathrm{AlN}$
waveguide and anneal the chip in an $\mathrm{O}_{2}$ atmosphere for
$\mathrm{5\,\mathrm{h}}$ at $\mathrm{950\textdegree C}$. The thickness,
width, and radius of $\mathrm{AlN}$ microring are $1\,\mathrm{\mu m}$,
$1.12\,\mathrm{\mu m}$ and $30\,\mathrm{\mu m}$, respectively. The
bus waveguide for IR light is $0.8\,\mathrm{\mu m}$ wide with a gap
to the microring of $0.6\,\mathrm{\mu m}$. The wrap around waveguide
for visible light is $80\,\mathrm{\mu m}$ long, tapered from $0.175\,\mathrm{\mu m}$
to $0.125\,\mathrm{\mu m}$, with a gap to the microring of $0.5\,\mathrm{\mu m}$.

\vbox{}

\noindent\textbf{Measurements. }Continuous-wave telecom laser (New
Focus 6427) is used as the drive laser. After amplification with an
EDFA, the drive laser is coupled into the device through an off-chip
wavelength division multiplexer (WDM) and a lensed fiber. An on-chip
WDM structure is used to guide the telecom drive laser into top bus
waveguide and form a counter-clockwise propagating field in the microring
resonator. In the NOIT experiment, we use a continuous-wave visible
laser (TLB-6712) to probe the transmission spectrum of the visible
mode. After coupling into the device through lensed fiber, the visible
probe laser is guided to the bottom wrap-around waveguide by the on-chip
WDM and forms either clockwise or counter-clockwise propagating light
field in the microring resonator. The transmitted visible light is
measured by a silicon photoreceiver (New Focus 2001). In frequency
conversion measurement, a second telecom laser (New Focus 6428) is
used to probe the high $Q$ resonance of mode $c$. The experimental
setup is shown in the supplementary information. 

\vbox{}

\noindent

\vbox{}

\noindent\textbf{Acknowledgments}\\ The authors thank Liang Jiang
for stimulating discussion and Michael Power and Dr. Michael Rooks
for assistance in device fabrication. H.X.T. acknowledges support
from DARPA and a Packard Fellowship in Science and Engineering. Facilities
used for device fabrication are supported by the Yale SEAS cleanroom
and the Yale Institute for Nanoscience and Quantum Engineering. 

\vbox{}

\noindent\textbf{Author contributions}\\ X.G. designed and fabricated
the samples, performed the measurements and analyzed the data, C.L.Z.
conceived the idea and provided theoretical analysis, H.J. assisted
the photonic chip fabrication. H.X.T. supervised the project. All
the authors discussed the results and X.G., C.L.Z. and H.X.T. wrote
the manuscript.

\vbox{}

\noindent\textbf{Competing financial interests}\\The authors declare
no competing financial interests.

\end{document}